\documentclass[12pt]{article}
\usepackage{graphicx}
\begin{document}
\markboth{Vladimir V. Belokurov Vsevolod V. Chistiakov Evgeniy T. Shavgulidze}{Perturbation Theory for Path Integrals in Quadratic Gravity}

%

%
\date{ }
\title{Perturbation Theory for Path Integrals in Quadratic Gravity}

\author{Vladimir V. Belokurov
\\{\small Lomonosov Moscow State University}\\
{\small vvbelokurov@yandex.ru}\\\\
Vsevolod V. Chistiakov\\
{\small Lomonosov Moscow State University}\\
{\small vsevolod.chistyakov@gmail.com}\\\\
Evgeniy T. Shavgulidze\\
{\small Lomonosov Moscow State University}\\
{\small shavgulidze@bk.ru}
}

\maketitle

\begin{abstract}
The action $A$ of Quadratic Gravity in FLRW metric is invariant under the group of diffeomorphisms of the time coordinate
and can be written in terms of the only dynamical variable $g(\tau)\,.$
We construct perturbation theory for calculating path integrals of the form $\int\,F(g)\,\exp\left\{-A (g)\right\}dg\,,$ and find the averaged value of the scale factor in the first nontrivial perturbative order.
\end{abstract}


\section{ Introduction}

\vspace{0.5cm}

Path integrals in quantum mechanics are considered as integrals over a Gaussian measure
$
\ \ \mu(dx)=\exp\left\{\imath\,A_{0}\right\}\, dx\ \
$
given by a free action $A_{0}\sim \int\ \frac{m\,(\dot{x})^{2}}{2}\,dt\,.$
{However, $\mu(dx) $ is not a countably-additive measure for functional integration.

In this case, Wick rotations are used for justification for Feynman path integrals. For imaginary time, the measure $\mu(dx)$ tansforms into the Wiener measure
$$
 w(dx)=\exp\left\{ -\frac{1}{2}\int \,\left(\dot{x}(t) \right)^{2}dt \right\}\ dx\,.
$$
}

All other terms in an action usually are considered as interaction terms. And if there is a small factor behind  them one can use perturbation theory calculating path integrals  by the well known rules.
The similar picture takes place  in the common models of fundamental interactions in QFT.

However there are actions of special types that lead directly to the Wiener measure after some substitutions of dynamical variables.

The non-linear non-local substitutions of the form
\begin{equation}\label{y x}
  y(t)=x(t)+\int \limits_{0}^{t}\,f\left(x(\tau)\right)\,d\tau\,
\end{equation}
realize the formal connection between the theories with interaction (including $x^{4}$ and Morse potentials) and the free one

$$w(dy)\equiv\exp\left\{ -\frac{1}{2}\int (\dot{y}(t))^{2}dt \right\}\ dy =$$
\begin{equation}\label{dydx}
 \exp\left\{ -\frac{1}{2}\int \left[\,f^{2}\left({x}(t)\right)-f'\left({x}(t)\right)\,\right]dt-BT\right\}\ w(dx)\,.
\end{equation}

{The equality of the measures can be easily proved in the discretized form of (\ref{y x}), interpreted in the Itô sense.}
Note that in (\ref{y x}) the integration spaces $X$ and $Y$ are different \cite{(BSh1)}, \cite{(BSh2)}.

Another example is given by the Schwarzian action\footnote{ It appears in the effective  theories for JT dilaton 2D gravity and SYK model
\cite{(SY)} - \cite{(Sar)}\, .}
$$
  A_{Sch}=-\frac{1}{\sigma^{2}}\int \limits _{[0,\,1]}\,\left[ \mathcal{S}ch (\varphi,\,t)+2\pi^{2}\left(\varphi'(t)\right)^{2}\right]dt\,,\ \ \ \
\mathcal{S}ch (\varphi,\,t)=
\left(\frac{\varphi''(t)}{\varphi'(t)}\right)'
-\frac{1}{2}\left(\frac{\varphi''(t)}{\varphi'(t)}\right)^2\,,
$$
The dynamical variable of the theory
 $\varphi(t)$ is an orientation preserving $(\varphi'(t)>0)$ diffeomorphism of the interval $
(\varphi\in Diff^{1}_{+}([0,\,1]))\,.$

The countably-additive  measure on the group of diffeomorphisms
$$
   \mu_{\mathcal{S}ch}(d\varphi)=\exp\left\{\frac{1}{\sigma^{2}}\int \limits _{[0,\,1]}\, \mathcal{S}ch (\varphi,\,t)\,dt  \right\}  d\varphi
$$
turns into the Wiener measure  $w_{\sigma}(d\xi)$ under the substitution of variables
$$
 \varphi(t)=\frac{\int \limits _{0}^{t}\,e^{\xi(\tau)}d\tau}{\int \limits _{0}^{1}\,e^{\xi(\eta)}d\eta }  \,,
\ \ \ \ \ \ \ \  \xi(t)=\log\varphi'(t)-\log\varphi'(0)\,,
$$
Here,
$\xi(t)$ is a continuous function on the interval $[0,\,1]$ satisfying the boundary condition
$\xi(0)=0\,.$

Thus the equality of functional integrals
$$
  \int\limits_{Diff^{1}_{+} ([0,\,1]) }F\left(\varphi,\,\varphi'\right)\,\mu_{\mathcal{S}ch}(d\varphi)
 =\int\limits _{C_{0}([0,1]) }F\left(\varphi(\xi),\,(\varphi(\xi))'\right)\, w_{\sigma}(d\xi)\,.
$$
gives us the possibility to calculate Schwarzian path integrals\,\, \cite{(BShExact)}, \cite{(BShCorrel)}, \cite{(BShCalc)}.

The idea to transform functional measure of a theory into the Wiener measure by some substitution seems particularly productive in complicated theories of quantum gravity. Quadratic gravity (see e.g. \cite{(AlvGaume)}-\cite{(Buchbinder)}) looks like a natural candidate for applying this approach.

In the recent paper\, \cite{(BSh22)}, we studied a
quadratic gravity model in FLRW metric. In this case, the action $A$ of the theory is invariant under the group of diffeomorphisms of the time coordinate, and can be written in terms of the only dynamical variable $g(\tau)\,$ that is diffeomorphism invariant.

{It is well known that, due to the fact that EH action is unbounded from below,
$$
\exp\left\{-A_{EH} \right\}\,d\mathcal{G}
$$
cannot be considered as a functional measure of Euclidean path integration.

To overcome the trouble with Euclidean path integrals in general relativity, several approaches were proposed.
(See, e.g.,
\cite{(Gibbons2)}-\cite{(Narain)}). The most radical ones avoid Wick rotation and study Feynman path  integrals directly.
The lack of a countably-additive measure for functional integration is a common feature of the approaches. Therefore they deal with the integrals that are not absolutely convergent.
Note that in the ordinary quantum field theory in the Minkowski space-time,
one has the opposite situation. Euclidean path integrals are used  as a justification for Feynman path integrals in the theory.

Another idea (\cite{(LoukoSorkin)}-\cite{(Visser)}) is to use the complex space-time metrics  to validate path integrals in gravity.

In\,\, \cite{(BSh22)} and in this paper, we propose the completely different approach to quantize gravity. We consider $R+R^{2}$ theory in the FLRW metric and find the dynamical variable $g(\tau)$ that is invariant under the group of diffeomorphisms of the time coordinate.
Then we turn  the Feynman path integrals
\begin{equation}
   \label{FPIA}
\int\,F(g)\,\exp\left\{i\,A (g)\right\}\,dg
\end{equation}
into the Euclidean ones
\begin{equation}
   \label{EPIA1}
\int\,F(g)\,\exp\left\{-A (g)\right\}\,dg
\end{equation}
by the corresponding transformation of the space-time metric.

Let us stress that we consider path integrals not over the space of metrics $\mathcal{G}$, as it is usually done, but
over the space of continuous functions $g(\tau)$ related to the conformal factor of the metric. It similar to  the idea
of\,\, \cite{(Antoniadis0)} $^{,}$  \cite{(Antoniadis2)} to consider the effective theory of the conformal factor as the true theory of quantum gravity.
 }

{In\, \cite{(BSh22)}\,, we proved the measure $\mu(g)=\exp\left\{-A(g) \right\}dg\,$  to be equivalent to the Wiener measure. We calculated  path integrals over the space of continuous functions with the initial conditions $g(0)=0\, \, ,g'(0)=0\, \,, g''(0)=const$ and found  perturbative corrections to the classical solution $a_{0}(t)= \sqrt{t}\,.$} 

{In this paper, we calculate path integrals over the space of continuous functions with the initial conditions $g(0)=0\,\, , g'(0)=1\,\, , g''(0)=1\,$ and get perturbative correction to the solution for the scale factor
exponentially  depending on time $a_0(t)=e^{t}\,.$} 

In section \ref{sec:qg}, we review briefly the main features of the quadratic gravity model studied and the results of \cite{(BSh22)} .

In section \ref{sec:solution}, we propose the non-linear non-local substitution that turns the functional measure for path integrals into the Wiener one and specify the classical solution with the scale factor
exponentially  depending on time.

Then, in section \ref{sec:pertcorr}, we calculate  the first nontrivial perturbative corrections to the averaged scale factor $<a(t)>$.

In section \ref{sec:concl}, we summarize our results and discuss briefly their possible physical interpretation.

Appendices  contain the details of calculations.

\section{ Quadratic Gravity model}
\label{sec:qg}

\vspace{0.5cm}

In the recent paper \cite{(BSh22)}, we studied the
quadratic gravity model with the action of the form
 $$
 \tilde{A} =\tilde{A}_{0}+\tilde{A}_{1}+\tilde{A}_{2}
$$
\begin{equation}
   \label{Aflrw}
 =\Lambda\int\,d^{4}x\,\sqrt{-\mathcal{G}}
-\frac{\kappa}{6}\,\int\,d^{4}x\,\sqrt{-\mathcal{G}}\,R+\frac{\lambda^{2}}{72}\int\,d^{4}x\,\sqrt{-\mathcal{G}}\,R^{2}
\end{equation}

\label{sec:inv}

\vspace{0.5cm}

in FLRW metric
$$
ds^{2} =N^{2}(\tilde{t})\,d\tilde{t}^{2}-a^{2}(\tilde{t})\,d\vec{x}^{2}\,,\ \ \ \ \ N(\tilde{t})>0\,,\ \ a(\tilde{t})>0\,.
$$

 {The general coordinate invariance of the action is reduced to its invariance under the group of reparametrizations of the time coordinate. We suppose it to be the group of diffeomorphisms of
 the real semiaxis including zero $ Diff \left( \mathbf{R}^{+}\right)$.

The two coordinate systems are the most popular. They are the so-called cosmological coordinate system where $N(t)=1\,,$ and the so-called conformal coordinate system where $N(\tau)=a(\tau)\,,$ with cosmological time $t$ and conformal time $\tau$ being the time variable in the corresponding coordinate system.

Consider  diffeomorphisms  $\varphi \in Diff \left( \mathbf{R}^{+}\right)\,.$
We define the action of the diffeomorphism $\varphi$ on the functions $N(\tilde{t})$ and $a(\tilde{t})$ as follows:
\begin{equation}
   \label{fiNa}
\varphi\,N(\tilde{t})=\left(\varphi^{-1}(\tilde{t}) \right)'\,N\left(\varphi^{-1}(\tilde{t}) \right)\,;\ \ \ \ \ \varphi\,a(\tilde{t})=a\left(\varphi^{-1}(\tilde{t}) \right)\,.
\end{equation}

Instead of the laps and the scale factors, it is convenient to use the functions $f(\tilde{t})$ and $h(\tilde{t})$ defined by the following equations:
\begin{equation}
   \label{f}
\left(f^{-1}(\tilde{t}) \right)'=\frac{N(\tilde{t})}{a(\tilde{t})}\,,\ \ \ \ \ \ \ f^{-1}(\tilde{t}) =\int\limits_{0}^{\tilde{t}}\,\frac{N(\tilde{t}_{1})}{a(\tilde{t}_{1})}\,d\tilde{t}_{1}\,,\ \ \ f^{-1}(0) =0\,;
\end{equation}
\begin{equation}
   \label{h}
h'(\tilde{t}) =N(\tilde{t})\,,\ \ \ \ \ \ \ h(\tilde{t}) =\int\limits_{0}^{\tilde{t}}\,N(\tilde{t}_{1})\,d\tilde{t}_{1}\,,\ \ \ h(0) =0\,,
\end{equation}
with the transformation rules\footnote{We denote the composition of the functions by "$\circ$ product". } under the action of the diffeomorphism $\varphi$
\begin{equation}
   \label{varphif-1}
(\varphi f)^{-1}(\tilde{t}) =\int\limits_{0}^{\tilde{t}}\,\frac{\varphi N(\tilde{t}_{1})}{\varphi a(\tilde{t}_{1})}\,d\tilde{t}_{1}=\int\limits_{0}^{\varphi^{-1}(\tilde{t})}\,\frac{N(\tilde{t}_{2})}{a(\tilde{t}_{2})}\,d\tilde{t}_{2}
=f^{-1}\left(\varphi^{-1}(\tilde{t})\right)\equiv
\left(f^{-1}\circ \varphi^{-1}\right)(\tilde{t}) \,;
\end{equation}
\begin{equation}
   \label{varphih}
(\varphi h)(\tilde{t}) =\int\limits_{0}^{\tilde{t}}\,\varphi N(\tilde{t}_{1})\,d\tilde{t}_{1}=\int\limits_{0}^{\varphi^{-1}(\tilde{t})}\,N(\tilde{t}_{2})\,d\tilde{t}_{2}=h\left(\varphi^{-1}(\tilde{t})\right)\equiv
\left(h\circ \varphi^{-1}\right)(\tilde{t}) \,.
\end{equation}

From (\ref{varphif-1}),  the equation
\begin{equation}
   \label{varphif}
\varphi f =
\varphi \circ f
\end{equation}
follows immediately.

The invariance of the action under the group of diffeomorphisms looks like
\begin{equation}
   \label{invA}
A(f,\,h)=A\left(\varphi f,\,\varphi h\right)=A\left(\varphi \circ f,\,h\circ\varphi ^{-1}\right)\,.
\end{equation}

Note that the function
\begin{equation}
   \label{barg}
 g= h\circ f
\end{equation}
 is invariant under the diffeomorphisms $\varphi\,:$
\begin{equation}
     g= h\circ \varphi^{-1} \circ \varphi \circ f=h\circ f\,.
\end{equation}

Choosing the diffeomorphism $\varphi$ to be
\begin{equation}
   \label{varphi=h}
\varphi  = h\,,
\end{equation}
we obtain the cosmological coordinate system
\begin{equation}
   \label{acosmol}
N(t)=1\,,\ \ \ \ \ a(t)=g'\left(g^{-1}(t) \right)\,,
\end{equation}
\begin{equation}
   \label{cosmolmetric}
ds^{2} =dt^{2}-\left(g'\left(g^{-1}(t) \right)\right)^{2}\,d\vec{x}^{2}\,.
\end{equation}

And if we choose,
\begin{equation}
   \label{varphi=f}
\varphi  = f^{-1}\,,
\end{equation}
we have the conformal coordinate system
\begin{equation}
   \label{confmetric}
ds^{2} =\left(g'(\tau)\right)^{2}\,\left[d\tau^{2}-d\vec{x}^{2}\right]\,,\ \ \ \ \ N(\tau)=a(\tau)=g'(\tau)\,.
\end{equation}

Therefore, the functions $f(\tilde{t})$ and $h(\tilde{t})$ have the transparent physical meanings.
$h$ is the reparametrization of the time $\tilde{t}$ into the cosmological time $t\,,$ while $f^{-1}$ stands for
the reparametrization of the time $\tilde{t}$ into the conformal time $\tau\,.$
}

The function
$$
 g(\tau)=\left( h\circ f\right)(\tau)=h\left(f(\tau)\right)
$$
transforms the conformal coordinate system into the cosmological one
\begin{equation}
   \label{taut}
t=g(\tau)\,,\ \ \ \ \ \ \ \ \ \  \tau=g^{-1}(t)\,.
\end{equation}

The invariance of the action manifests itself in its dependence on the only invariant function $g$
\begin{equation}
   \label{AgRight}
A=A\left(g\right)=A_{0}\left(g\right)+A_{1}\left(g\right)+A_{2}\left(g\right)\,,
\end{equation}
with the explicit form
\begin{equation}
   \label{Ag0}
A_{0}\left( g\right)=\Lambda\,\int\,\left(g'(\tau) \right)^{4}\,d\tau\,,
\end{equation}
\begin{equation}
   \label{Ag1}
A_{1}\left( g\right)=-\kappa\,\int\,\left[\left(g''(\tau) \right)^{2}-\left(g''(\tau)\,g'(\tau) \right)'\,\right]\,d\tau\,,
\end{equation}
and
\begin{equation}
   \label{Ag2}
A_{2}\left( g\right)=\frac{\lambda^{2}}{2}\,\int\,\left(\frac{g'''(\tau)}{ g'(\tau)} \right)^{2}\,d\tau\,.
\end{equation}

{Note that from (\ref{f}) - (\ref{barg}), it follows that for the substitutions $\{N\,,\ a\} \ \rightarrow \  \{f\,,\ h \}  \ \rightarrow \  \{g\,,\ \varphi \}$

   $$\frac{\int \mathcal{F}(N,a) \exp\left\{-A (N,\,a)\right\}\,dN\,da\,}{\int \exp\left\{-A (N,\,a)\right\} dN da} = \frac{\int \,\tilde {\mathcal{F}}(f,h) \exp\left\{-A (f,\,h)\right\}\,df\,dh\,}{\int \exp\left\{-A (f,\,h)\right\} \,df\,dh}=$$
   
\begin{equation}
   \label{Meas} \frac{\int F(g)\exp\left\{-A (g)\right\}\,dg\,d\varphi\,}{\int \exp\left\{-A (g)\right\}\,dg\,d\varphi\,}=\frac{\int \,F(g) \exp\left\{-A (g)\right\}\,dg\,}{\int \exp\left\{-A (g)\right\} dg}.
 \end{equation}}

{It is a unique example of a theory where the dynamical variable is gauge invariant and there is the direct product of the measures corresponding to the dynamical variable and to the gauge degree of freedom.}

{It is completely different from the Faddeev-Popov approach to quantization
where dynamical variables are taken in a fixed gauge and ghosts are needed. }

\vspace{0.5cm}

\section{The classical solution and the non-linear non-local substitution}
\label{sec:solution}

\vspace{0.5cm}

Henceforth we  consider the model with the particular relation between the parameters
\begin{equation}
\label{param_cond}
    \lambda=\frac{\kappa}{3\sqrt{2\Lambda}}\,,
\end{equation}
and write down the action of the quadratic gravity model in the form 
\begin{equation}
   \label{Anew}
A=\frac{\lambda^2}{2}\,\int_0^{1} d\tau \left[\left(\frac{g''}{g'}\right)'-\left(\frac{g''}{g'}\right)^2+\frac{\kappa}{3\lambda^2}(g')^2\right]^2\,.
\end{equation}
{It differs from (\ref{AgRight}) by the boundary term
$$
-\frac{2}{3}\int d\tau \,
\left(\kappa\,g''\,g'+\lambda^2 \left[\frac{g''}{g'}\right]^3\right)'\,.
$$
}

{Note that the cosmological time $t\in [0,\infty)$, and the conformal one $\tau\in [0,1]$ (see (\ref{tau pert}) below )\,.}

To calculate path integrals in the theory it is convinient to perform the non-linear non-local substitution
  $p\left(g(\tau)\right)$
\begin{equation}
\label{g p}
\left(\frac{g''}{g'}\right)'-\left(\frac{g''}{g'}\right)^2+\frac{\kappa}{3\lambda^2}(g')^2=\frac{1}{\lambda}p'(\tau)\,.
\end{equation}

{Eq. (\ref{g p}) can be regarded as the two-step substitution. On the first step
$$\beta(\tau)=\frac{g''(\tau)}{g'(\tau)}\, , \qquad g(\tau)=\int_0^{\tau}d\tau_1 e^{\int_0^{\tau_1} d\tau_2 \, \beta(\tau_2)}\, \, .$$
As it was shown in\,\, \cite{(Shavgulidze1988)}, \cite{(Shavgulidze2000)}, \cite{(BShCalc)}, the following equality for the normalized measures is valid
$$\exp(-A(g))dg=\exp\left(-A(g(\beta))\right)d\beta\,.$$
On the second step, the substitution
$$\frac{1}{\lambda}p(\tau)=\beta(\tau)-\int_0^{\tau} d\tau_1 \left[ \beta^2(\tau_1)-\frac{\kappa}{3\lambda^2}e^{2\int_0^{\tau_1} d \tau_2 \beta(\tau_2) }\right]$$ is of the same type as (\ref{y x}).
}
Now the functional measure of the theory turns into the Wiener measure
\begin{equation}
    \mu(dg)\equiv\exp(-A(g))dg=\exp\left(-\int_0^{1}\frac{(p'(\tau))^2}{2}d\tau\right)dp\equiv w(dp)\,.
\end{equation}

{The functional $-\int_{0}^1 d\tau (p'(\tau))^2/2$ defined on the space of functions $p(\tau)$, with the condition $p(0)=0$ and $p(1)$ not fixed, has a unique maximum at $p'(\tau)=0$. We integrate over functions with the initial conditions $g(0)=0\,\,, g'(0)=1\,\,, g''(0)=1$. So, there is the one to one correspondence between the functions $g(\tau)$ and $p(\tau)$.}
The maximal input into functional integrals (\ref{EPIA1}) is provided by the paths with vanishing {right-hand side of the (\ref{g p})}
\begin{equation}
\label{gcl}
    \left(\frac{g''}{g'}\right)'-\left(\frac{g''}{g'}\right)^2+\frac{\kappa}{3\lambda^2}(g')^2=0\,.
\end{equation}

Equation (\ref{gcl}) is equivalent to the system
\begin{equation}
\label{qhcl}
\left\{ \begin{array}{c}
{\beta}'={\beta}^2-\frac{\kappa}{3\lambda^2} {\alpha}^2\\
{\alpha}'={\beta \alpha}
\end{array}\right.
\end{equation}
where
$${\beta}\equiv\frac{g''}{g'} \,,\ \ \ \ \  {\alpha}\equiv g'\,.$$

Representing ${\beta}(\tau)$ as ${\beta}(\tau)={\beta}({\alpha}(\tau))$ we get linear equation for  ${\beta}^2$
$$
\frac{d{\beta}^2}{d{\alpha}}=\frac{2{\beta}^2}{{\alpha}}-\frac{2\kappa}{3\lambda^2}{\alpha}
$$
$$
{\beta}^2({\alpha})={ \alpha}^2\left(1-\frac{2\kappa}{3\lambda^2}\ln {\alpha}\right)\,.
$$
The solution ${\alpha}(\tau)$ is given by the implicit function
\begin{equation}
\int_1^{{\alpha}(\tau)} \frac{dy}{y^2\sqrt{1-\frac{2\kappa}{3\lambda^2}\ln y}}=\tau\,.
\end{equation}

For $g(\tau)=g({\alpha}(\tau))\,,\ $ and $ \frac{dg}{d{\alpha}}=\frac{1}{{\beta}({\alpha})}\,,$
we get
\begin{equation}
    g(\tau)=-\frac{3\lambda^2}{\kappa}\left(\sqrt{1-\frac{2\kappa}{3\lambda^2}\ln {\alpha}(\tau)}-1\right)\,,
\end{equation}
and
\begin{equation}
\label{h(g)}
    {\alpha}(\tau)=\exp\left(g(\tau)-\frac{\kappa}{6\lambda^2}g^2(\tau)\right)\,.
\end{equation}

From (\ref{h(g)}) it follows that
\begin{equation}
    \tau(t)=\int_0^{t}dy\, \exp\left(-y+\frac{\kappa}{6\lambda^2}y^2\right)\,.
\end{equation}
It can be considered as the implicit solution for $g(\tau)\,$
\begin{equation}
\label{g(tau)}
    \tau=\int_0^{g(\tau)}dy\, \exp\left(-y+\frac{\kappa}{6\lambda^2}y^2\right)\,.
    \end{equation}
Expanding the integral in the series we have
\begin{equation}\label{tau pert}
    \tau(t)=g^{-1}(t)=1-e^{-t}+O(\lambda^{-2})\,,
\end{equation}
and
\begin{equation}
    g(\tau)=-\ln(1-\tau)+O(\lambda^{-2})\,.
\end{equation}

Taking into account the relations between the functions ${\beta}(\tau)$, ${\alpha}(\tau)$ and the scale factor $a(t)$
\begin{equation}
    {\alpha}(\tau)=a(g(\tau))\, , \qquad {\alpha}'=\frac{da}{dt}\frac{dt}{d\tau}=\dot a a\,,
\end{equation}
\begin{equation}
    {\beta}=\dot a\, , \qquad {\beta}'=\ddot aa\, ,
\end{equation}
we have the equation
\begin{equation}
\label{aeq}
   \ddot aa=\dot a^2 -\frac{\kappa}{3\lambda^2}a^2\,.
\end{equation}
For Hubble parameter $H=\frac{\dot a}{a}$ equation (\ref{aeq}) looks like
\begin{equation}
    \dot H=-\frac{\kappa}{3\lambda^2}\,,
\end{equation}
with the solution
\begin{equation}
   H(t)=1-\frac{\kappa}{3\lambda^2}t\, ,\qquad a(t)=\exp\left(t-\frac{\kappa}{6\lambda^2}t^2\right)\,.
\end{equation}

At the moment $t_m=\frac{3\lambda^2}{\kappa}$ the scale factor has the maximal value $a_m=\exp\left(\frac{3\lambda^2}{2\kappa}\right)$
and the solution can be rewritten as
\begin{equation}
   a(t)=a_m\exp\left(-\frac{(t-t_m)^2}{2t_m}\right)\, , \qquad H(t)=1-\frac{t}{t_m}\,.
\end{equation}
Note that the solution is singular free (cf.\cite{Starob}).

And the curvature has the form
\begin{equation}
    R=-6\left(\frac{\ddot a}{a}+\frac{\dot a^2}{a^2}\right)=-6\left(\dot H+2H^2\right)=\frac{2\kappa}{\lambda^2}-12\left(1-\frac{t}{t_m}\right)^2\,.
\end{equation}

In the next section, we solve (\ref{g p}) perturbatively and calculate the scale factor up to the perturbative order $\lambda^{-2}\,.$

 \vspace{0.5cm}

\section{First nontrivial perturbative correction to the averaged scale factor}
\label{sec:pertcorr}

\vspace{0.5cm}

To calculate the averaged value of scale factor
\begin{equation}
    \langle a(t)\rangle \equiv \int g'(g^{-1}(t)) \,w(dp)
\end{equation}
first we represent the solution $g(p(\tau))$ of (\ref{g p}) as the series
$$
g(\tau,\lambda)=\sum_k g_k(\tau)\lambda^{-k}\,.
$$
It is convenient to solve pertubatively  the system equivalent to (\ref{g p})
\begin{equation}
\label{qh}
\left\{ \begin{array}{c}
{\beta}'={\beta}^2-\frac{\kappa}{3\lambda^2} {\alpha}^2+\frac{1}{\lambda}p'(\tau)\\
{\alpha}'={\beta}{\alpha}
\end{array}\right.
\end{equation}
where
$$
{\beta}(\tau)=\sum_{k}\frac{{\beta}_k(\tau)}{\lambda^k} \,,\ \ \ \ \  {\alpha}(\tau)=\sum_{k} \frac{{\alpha}_k(\tau)}{\lambda^k}\,.
$$
Up to the second order we have
\begin{equation}\label{qh0}
\left\{ \begin{array}{c}
{\beta}'_0={\beta}_0^2\\
{\alpha}'_0={\beta}_0{\alpha}_0 \, ,
\end{array}\right.
\end{equation}

\begin{equation}\label{qh1}
\left\{ \begin{array}{c}
{\beta}'_1=\frac{2}{1-\tau}{\beta}_1+p'(\tau)\\
{\alpha}'_1=\frac{1}{1-\tau}{\beta}_1+\frac{1}{1-\tau}{\alpha}_1 \, ,
\end{array}\right.
\end{equation}

\begin{equation}\label{qh2}
\left\{ \begin{array}{c}
{\beta}'_2=\frac{2}{1-\tau}{\beta}_2+{\beta}_1^2-\frac{\kappa}{3} \left(\frac{1}{1-\tau}\right)^2\\
{\alpha}'_2=\frac{1}{1-\tau}{\beta}_2+\frac{1}{1-\tau}{\alpha}_2+{\alpha}_1{\beta}_1\, ,
\end{array}\right.
\end{equation}

The solutions of the systems are given in Appendix A.

At the next step we represent the coefficients $a_k$ of the series
\begin{equation}
    a(t,\lambda)=g'(\tau(t,\lambda),\lambda)=\sum_k a_k(t)\lambda^{-k}
\end{equation}
as functions of $g_k(\tau)\,$ where
$$
\tau(t,\lambda)=\sum_k \tau_k(t)\lambda^{-k}
$$
 is the solution of the equation
\begin{equation}
\label{gt}
g(\tau(t,\lambda),\lambda)=t\,.
\end{equation}
As the result, we find (Appendix B) the coefficients
 $a_k(t)\,$
\begin{equation}\label{a0simp}
    a_0(t)=e^{t}\,,
\end{equation}
\begin{equation}\label{a1simp}
    a_1(t)=- g_{1} e^{t} + {\alpha}_{1}\,,
\end{equation}
\begin{equation}\label{a2simp}
    a_2(t)=\frac{g_{1}^{2} e^{t}}{2} - g_{1} {\beta}_{1} - g_{2} e^{t} + {\alpha}_{2}
\end{equation}
where
${\beta}_k={\beta}_k(\tau_0(t))\,,\ \ \ {\alpha}_k={\alpha}_k(\tau_0(t))\,,\ \ \ g_k=g_k(\tau_0(t))$ are the solutions of (\ref{qh0})-(\ref{qh2}) at $\tau_0(t)=1-e^{-t}\,.$

Path integration (Appendix C) leads to

\begin{equation}\label{a2res}
    \langle a_1(t)\rangle=0\,,\ \ \ \ \
\langle a_2(t)\rangle = \left((1-\kappa)\frac{t^{2}}{6} - \frac{t}{9} + \frac{1}{27}\right) e^{t} - \frac{e^{- 2 t}}{27}\,.
\end{equation}
Thus up to the second order, the averaged scale factor has the form
\begin{equation}\label{ares}
\langle a(t)\rangle = \left(1+\frac{1}{\lambda^2}\left[(1-\kappa)\frac{t^{2}}{6} - \frac{t}{9} + \frac{1}{27}\right]\right) e^{t} - \frac{e^{- 2 t}}{27\lambda^2}\,.
\end{equation}

\vspace{0.5cm}

\section{Conclusion}
\label{sec:concl}

\vspace{0.5cm}

In this paper, we study path integrals in the quadratic gravity model in the FLRW metric. The general coordinate invariance of the theory is reduced in this case to its invariance under the group of diffeomorphisms of the time coordinate. The invariant of the group $g(\tau)$ can be considered as the only dynamical variable in the theory. Thus we deal with the well-defined path integrals avoiding any ghosts problems.

In our approach,  Euclidean path integrals
are functional integrals not over the space of metrics $\mathcal{G}$, as they are commonly defined, but functional integrals
over the space of continuous functions $g(\tau)$
$$
\int\,F(g)\,\mu(dg)\,,\ \ \ \ \
\mu(dg)=\exp\left\{-A \right\}dg\,.
$$

We prove the measure  $\mu(g)$ to be equivalent to the Wiener measure and therefore not only rigorously define path integrals in quadratic gravity but also give the regular method of calculation. As an example of the approach, we calculate the first nontrivial perturbative  correction to the averaged scale factor $<a(t)>\,.$

In the previous paper \cite{(BSh22)}, we studied perturbative corrections to the scale factor for
 the classical solution that gives the model of the universe arising from a point and expanding with the slowdown \cite{GorbRub1}.

 Here, we consider another classical solution corresponding to the inflationary phase of the universe evolution \cite{GorbRub2} {and calculate corrections to the averaged scale in the first nontrivial order} .

\vspace{0.5cm}

\section{Acknowledgements}

\vspace{0.5cm}

We would like to express our deep gratitude to a prominent person and a brilliant scientist - Valeriy Rubakov who suddenly passed away two years ago.
For many years VVB and ETSh had the good fortune to make use of amicable criticism, valuable advices and steady moral support from our dear friend.

\vspace{0.5cm}

\section{Appendix A}

\vspace{0.5cm}

Here, we write down the solutions for the coefficients ${\beta}_{k}$ and ${\alpha}_{k}\,,\ \ \ \ k=0,\,1,\,2.$

$${\beta}_0={\alpha}_0=\frac{1}{1-\tau}\,,$$

\begin{equation}\label{q1sol}
    {\beta}_1(\tau)=\int_0^\tau ds \frac{(1-s)^2}{(1-\tau)^2}p'(s)= p(\tau)+\frac{2}{(1-\tau)^2}\int_0^\tau ds (1-s) p(s)\,,
\end{equation}

\begin{equation}\label{h1sol}
    {\alpha}_1(\tau)=\frac{1}{1-\tau}\int_0^{\tau}ds\,\,\, {\beta}_1(s)\, ,
\end{equation}

\begin{equation}\label{g1sol}
    g_1(\tau)=\int_0^{\tau}ds\,\, L_g(\tau,s){\beta}_1(s)\, , \qquad L_g(\tau,s)=\ln\left(\frac{1-s}{1-\tau}\right)\, ,
\end{equation}

\begin{equation}\label{q2sol}
   {\beta}_2(\tau)=\frac{1}{(1-\tau)^2}\left\{\int_0^{\tau} ds \, (1-s)^2 {\beta}_1^2(s)+2(1-\tau)\int_0^{\tau} ds {\beta}_1(s) \right\}\, ,
\end{equation}

\begin{equation}\label{h2sol}
    {\alpha}_2(\tau) = \int_0^{\tau} ds K_{hq}(\tau,s) \left\{{\beta}_1^2(s)-\frac{\kappa}{3(1-s)^2} \right\}+K_{hh}(\tau,s){\alpha}_1(s){\beta}_1(s)\, ,
\end{equation}

\begin{equation}\label{g2sol}
g_2(\tau) = \int_0^{\tau}ds K_{gq}(\tau,s) \left\{{\beta}_1^2(s)-\frac{\kappa}{3(1-s)^2} \right\}+K_{gh}(\tau,s){\alpha}_1(s){\beta}_1(s)\, ,
\end{equation}

\vspace{0.5cm}

where we use the following definitions:
$$K_{hq}=\frac{(1-s)^2}{(1-\tau)^2}-\frac{1-s}{1-\tau}\, , \qquad K_{hh}=\frac{1-s}{1-\tau}\, ,$$
$$K_{gq}=\frac{(1-s)^2}{1-\tau}-(1-s)-(1-s)\ln\left(\frac{1-s}{1-\tau}\right)\, , \qquad K_{gh}=(1-s)\ln\left(\frac{1-s}{1-\tau}\right)\, .$$

\vspace{0.5cm}

\section{Appendix B }

\vspace{0.5cm}

Consider a composition of arbitrary functions ${\psi}$ and $\tau$ depending on parameter $\lambda$
\begin{equation}
\chi(t,\lambda)={\psi}(\tau(t,\lambda),\lambda).
\end{equation}

Coefficients ${\psi}_k$ of the series
\begin{equation}
{\psi}(\tau,\lambda)=\sum_{k=0}^{\infty}{\psi}_k(\tau)\lambda^{-k}
\end{equation}
can be represented as

\begin{equation}
{\psi}_k(\tau)=\sum_{s=0}^{\infty}{\psi}_{ks}(\tau-\tau_0)^s, \qquad {\psi}_{ks}(t)=\frac{1}{s!}\frac{d^s{\psi}_k}{d \tau^s}|_{\tau=\tau_0(t)}\,\, .
\end{equation}

To calculate the coefficients $\chi_k(t)$ of the expansion:
\begin{equation}
\chi(t,\lambda)={\psi}(\tau(t,\lambda),\lambda)=\sum_{k=0}^{\infty}\chi_k(t)\lambda^{-k}
\end{equation}
we define the $T^{(s)}_{m}$ as:

\begin{equation}
(\tau-\tau_0)^s=\left(\sum_{k=1}^{\infty}\tau_k\lambda^{-k}\right)^s=\sum_{m=0}^{\infty} T_m^{(s)} \lambda^{-m}.
\end{equation}

In this case, the recurrent relation
\begin{equation}
T^{(s)}_m=\sum_{r=s-1}^{m-1} T^{(s-1)}_r\tau_{m-r}
\end{equation}
is valid.

Now there is the equation for $\chi_k$
\begin{equation}\label{answ}
\chi_k(t)=\sum_{l=0}^{k} \sum_{s=0}^{k-l}{\psi}_{ls}T^{(s)}_{k-l} \,\, .
\end{equation}

Eq. (\ref{answ}) can be considered as the system for $\tau_k$

\begin{equation}
\chi_0=g_0(\tau_0(t))\,\, , \qquad \chi_1=g_{00}T^{0}_1+g_{01}T^{1}_1+g_{10}T^{0}_0=g'_0(\tau_0(t))\tau_1(t)+g_1(\tau_0) \,\, ,
\end{equation}

\begin{equation}
\chi_2=g_{20}+g_{01}\tau_2+g_{11}\tau_1+g_{02}\tau_1^2 \,\, ,
\end{equation}

\begin{equation}
t=g_0(\tau_0(t)) \Longrightarrow \tau_0(t)=g_0^{-1}(t)\,\, , \qquad \chi_1=0  \Longrightarrow \tau_1(t)=-\frac{g_1(\tau_0(t))}{g'_0(\tau_0(t))}=-\frac{g_{10}}{g_{01}}\,\, ,
\end{equation}

\begin{equation}
\chi_2=0 \Longrightarrow \tau_2=-\frac{g_{20}}{g_{01}}+\frac{g_{11}g_{10}}{g_{01}^2}-\frac{g_{02}g_{10}^2}{g^3_{01}}\,\, .
\end{equation}

To find $a_k$ we substitute $\tau_k$ into (\ref{answ}) with $f=g'$
\begin{equation}\label{a0}
a_0=g_{01}\,\, ,
\end{equation}
\begin{equation}\label{a1}
a_1=g_{11} - \frac{2 g_{02} g_{10}}{g_{01}}\,\, ,
\end{equation}

\begin{equation}\label{a2}
a_2=g_{21} - \frac{2 g_{02} g_{20}}{g_{01}} - \frac{2 g_{10} g_{12}}{g_{01}} + \frac{2 g_{02} g_{10} g_{11}}{g_{01}^{2}} + \frac{3 g_{03} g_{10}^{2}}{g_{01}^{2}} - \frac{2 g_{02}^{2} g_{10}^{2}}{g_{01}^{3}}\,\, .
\end{equation}

Taking into account the equations
\begin{equation}\label{gij}
g_{0n}=\frac{1}{n}e^{nt}\,\, , \qquad g_{n0}=g_n\,\, , \qquad g_{n1}={\alpha}_n\,\, , \qquad  g_{12}=\frac{{\alpha}_{1} e^{t}}{2} + \frac{{\beta}_{1} e^{t}}{2}
\end{equation}

we obtain (\ref{a0simp})-(\ref{a2simp}).

\vspace{0.5cm}

\section{Appendix C}

\vspace{0.5cm}

We represent $\langle a_2(t)\rangle$ in the following form

$$\langle a_2(t)\rangle=\frac{J_{1} e^{t}}{2} - J_{2} - J_{3} e^{t} + J_{4}$$
where
$$J_1=\langle g_1^2 \rangle\, , \qquad J_2=\langle g_1{\beta}_1 \rangle\, , \qquad J_3=\langle g_2 \rangle\, , \qquad J_4=\langle {\alpha}_2 \rangle\, .$$

It's convenient to use the auxiliary averages

$$C(s,s_1)\equiv \langle {\beta}_1(s){\beta}_1(s_1)\,  \rangle=\frac{1-(1-s)^5}{5(1-s)^2(1-s_1)^2}, \qquad (s\le s_1)\, ,$$

$$R(s)\equiv\langle {\beta}_1(s){\alpha}_1(s) \rangle =\frac{1}{5(1-s)}\left(\frac{1}{(1-s)^3}+\frac{(1-s)^2}{4}-\frac{5}{4(1-s)^2}\right)\, .$$

From (\ref{g1sol})-(\ref{g2sol}) we obtain

$$J_1=\int_0^{\tau} ds \int_0^s ds_1 F_1(\tau,s,s_1)\,\, , \qquad F_1(\tau,s,s_1)=2L_g(\tau,s)L_g(\tau,s_1)C(s_1,s)\,\, ,$$

$$J_2=\int_0^{\tau}ds F_2(\tau,s)\,\, , \qquad F_2(\tau,s)=L_g(\tau,s)C(s,\tau)\,\, , $$

$$J_3=\int_0^{\tau} ds F_3(\tau,s)\,\, , \qquad F_3(\tau,s)=K_{gq}(\tau,s)\left\{C(s,s)-\frac{ \kappa}{3(1-s)^2}\right\}+K_{gh}(\tau,s)R(s)\,\, ,$$

$$J_4=\int_0^{\tau} ds F_4(\tau,s)\,\, , \qquad F_4(\tau,s)=K_{hq}(\tau,s)\left\{C(s,s)-\frac{ \kappa}{3(1-s)^2}\right\}+K_{hh}(\tau,s)R(s)\,\, .$$

Combining the results of integration

$$J_{1}=\frac{t^{2}}{3} - \frac{t e^{t}}{2} + \frac{4 t}{9} + \frac{e^{2 t}}{5} - \frac{3 e^{t}}{8} + \frac{5}{27} - \frac{11 e^{- 3 t}}{1080}\,,$$
$$J_{2}=- \frac{t e^{2 t}}{4} + \frac{e^{3 t}}{5} - \frac{3 e^{2 t}}{16} - \frac{e^{- 2 t}}{80}\,,$$
$$J_{3}=\frac{\kappa}{3}\left(\frac{t^{2}}{2} +  t - e^{t} + 1\right) + \frac{t}{3} + \frac{e^{2 t}}{10} - \frac{e^{t}}{2} + \frac{7}{18} + \frac{e^{- 3 t}}{90}\,,$$
$$J_{4}=\frac{\kappa}{3}\left( t e^{t} -  e^{2 t} +  e^{t}\right) + \frac{e^{3 t}}{5} - \frac{e^{2 t}}{2} + \frac{e^{t}}{3} - \frac{e^{- 2 t}}{30}$$

we get (\ref{a2res}), (\ref{ares}).

\end{document}